# Developing and Evaluating Lightweight Cryptographic Algorithms for Secure Embedded Systems in IoT Devices

Brahim Khalil Sedraoui[1], Abdelmadjid Benmachiche[1], Amina Makhlouf[1]

[1] *University of Chadli Bendjedid, Faculty of Sciences & Technology, El Tarf, Algeria*

**Abstract**

The high rate of development of Internet of Things (IoT) devices has brought to attention new challenges in the area of data security, especially within the resource-limited realm of RFID tags, sensors, and embedded systems. Traditional cryptographic implementations can be of inappropriate computational complexity and energy usage and hence are not suitable on these platforms. This paper examines the design, implementation, and testing of lightweight cryptographic algorithms that have been specifically designed to be used in secure embedded systems. A comparison of some of the state-of-the-art lightweight encryption algorithms, that is PRESENT, SPECK, and SIMON, focuses on the main performance indicators, i.e., throughput, use of memory, and energy utilization. The study presents novel lightweight algorithms that are founded upon the Feistel-network architecture and their safety under cryptanalytic attacks, e.g., differential and linear cryptanalysis. The proposed solutions are proven through hardware implementation on the FPGA platform. The results have shown that lightweight cryptography is an effective strategy that could be used to establish security and maintain performance in the IoT and other resource-limited settings.

## 1. Introduction

The rapid development of IoT technologies has allowed many low-cost sensor-based devices to be widely used in smart homes and other connected environments [1], including digital learning ecosystems such as MOOCs [2], [3], [4] and Open Classrooms [5], [6], [7], and has accelerated the adoption of this technology; however, it has also raised tremendous challenges in terms of security and privacy [8]. Given that users may have privacy concerns on sharing identity-revealing data with the service providers, there is a requirement for mechanism of privacy preserving data aggregation to enable analytical operation without/with minimum disclosure of personalized information. To solve this, a smart home architecture is presented that integrates distributed smart sensors with a central hub for data filtering and processing, supporting secure analytics while enabling intrusion detection [9], phishing detection, and behavior-level cheating detection [10] as part of a broader IoT security posture as well as countermeasures of hardware efficient light weight cryptography for IoT environments.

IoT solutions are physical devices mixed with data-driven parts placed within a larger ICT infrastructure. Their deployment in safety-critical or data-privacy-sensitive applications makes them vulnerable to cyberattacks, privacy information leakage and large-scale exploits. Due to resource limitations of IoT devices and the Sensitive nature of data exchange, study stresses on lightweight cryptographic methods in order to improve security, protect user's privacy and assist secure deployment of large-scale IoT and smart home [11].

## 2. Fundamentals of Cryptography

Cryptography is the field of protecting information, such as text, sound or pictures, by using encryption and decryption. It is done based on cryptographic keys to convert understandable plaintext to unreadable form and vice versa. Cryptography consists of symmetric and asymmetric. Symmetric encryption uses a key to encrypt and decrypt data, including block ciphers (block of fixed size, e.g., DES/AES/SEED) and stream ciphers (encrypting bits/bytes in sequence), which is preferable in limited environment (e.g., appliances) or continuous transmitting environments. Asymmetric Cryptography (Public-key Cryptography) Asymmetrical cryptography, or public key cryptography, involves a pair of keys: one to encrypt data and the other to decrypt. Though it does secure key distribution without sharing secrets, it is computationally intensive and often exclusively applied to secure symmetric keys [12].

Now-a-days ciphers are used which work on input data of fixed size, such kind of function requires padding as well as if block sizes are small then they have to suffer from certain attacks. Unlike block ciphers, stream ciphers encrypt data one byte at a time allowing for slow (even real-time) encryption of transmitted messages that are meant to be secure but do not require protection beyond what is offered by the underlying cryptographic algorithm [13].

## 3. Lightweight Cryptography

Lightweight cryptography refers to the design of cryptographic primitives which minimize computation, memory usage and energy consumption so that they could be implemented on the ultra-restrictive resource devices, e.g. microcontrollers, sensor nodes and RFID tags. Contrasting with classical cryptography, lightweight solutions trade-off security and efficiency and can be used to meet the severe resource constraints in IoT devices that are found everywhere including smart homes, healthcare, industry, transportation etc [14].

The IoT devices in particular the resource-constraint devices, e.g., the passive tags of the RFID technology are susceptible to attacks like eavesdropping and data tampering. Therefore, it is required that cryptographic primitives should be designed specifically for such underlaid platforms. Larger capacity devices such as drones or cameras have larger data rates requiring less compute-intensive techniques that are secure enough but yet efficient to be used in practice [15].

### 3.1. Criteria for Lightweight Algorithms

Include low memory and energy requirements, platform independence, and suitability for small data sizes. They are essential for ensuring confidentiality and authenticity in secure embedded systems [16], wireless sensor networks, and other IoT applications. The section also highlights emerging lightweight block and stream ciphers, emphasizing their design considerations and challenges in implementation and practical deployment.

## 4. Existing Lightweight Cryptographic Algorithms

This field of lightweight ciphers for embedded and IoT devices has been studied with a focus on the efficiency, compactness and security. Therein, algorithms are to protect data on devices with weak computation capability, limited memory space and energy resources in the lightest way possible—in hardware abstraction level—having their implementations tested on FPGA platforms were using tools such as Intel RAPL for power consumption calculation is routine.

- **PRESENT:** a lightweight block cipher with 64-bit block size specifically designed for small area and inexpensive hardware implementation as required by RFID tags and sensors [11].
- **SPECK:** a lightweight ISO-standardized small block cipher designed for embedded systems and IoTs that can be accommodated in miniaturized SoC designs and implemented on the devices at-price to allow secure AI based edge processing to run in low-resource environments [17].
- **SIMON:** developed by the NSA in 2013, is a family of lightweight block ciphers with multiple choice for block and key size (48–256 bits) targeting low-energy environments. Its round function is simple —consisting of three 'and', one 'xor' and two rotations— allowing it to work efficiently in hardware, and be applicable to diverse IoT systems [18].

In all, these algorithms illustrate the trade-offs among compactness, efficiency and security, thus yielding actionable answers for securing constrained embedded systems and IoTs.

## 5. Challenges in Lightweight Cryptography

The designing lightweight cryptographic algorithms has to address serious issues which include the trade-offs between the security and performance. In general, stronger algorithms require a higher level of memory or computation power (computation complexity), which does not always meet the specifications for the resource-constrained devices. Designers need to prepare for new miniature hardware-enabled attack surfaces and anticipate wider implementation decisions (hardware, software or mixed) to ensure future-proof security [19].

Energy efficiency is also an important issue, especially for IoT devices running on batteries. Precise power measurement utilizing instruments like JouleScope or EnergyTrace™ is necessary to analyze forward and reverse performance, energy per processed bit and total energy consumption in order to make sure that light weight schemes can satisfy security goals and requirements in resource limited environments [20].

## 6. Research Methodology

The approach followed in the study is a step-by-step method for developing and applying lightweight cryptographic algorithms on IoT-embedded devices. Firstly, three Feistel-based lightweight algorithms on the security after performance such as differential and linear analysis, area efficiency and processing speed are presented. Secondly, an FPGA-based system-on-chip (SoC) is designed, in which the processor and specific peripherals for encryption, authentication and key generation are included to perform secure data processing. Third, hardware analysis is performed on Xilinx Virtex 5 ML501 FPGAs which basically assesses access control, data confidentiality and system operation by means of logic analyzer and simulation testing. Fourthly, hardware performance indicators, e.g., resource usage, operating frequency, power and energy per bit and throughput are evaluated to verify if resource-limited IoT devices can execute the scheme. Performance is measured quantitatively For computational speed there is a formula for the same: [11]

$$Throughput\ (bits/second) = N\ /\ T$$

Where:
- $N$ is the number of plaintext bits.
- $T$ is the time taken in seconds for encryption

Memory requirement is also efficient for the embedded environment such as 61.7 bytes are taken for 3 rounds of scr-CMAC; hence memory storage is expensive but our schemes show that they provide high computational overhead to keep them secure. This approach makes it possible to be sure about the practicality and security of cryptographic proposals for embedded/IoT applications.

# 7. Case Studies and Applications

The usage of lightweight cryptographic algorithms to secure IoT environments such as smart grids and embedded platforms is illustrated through case studies in this section. These architectures are designed taking account of the low computational ability and to maintain secure data exchange between modules as well as to counteract attacks which may include spreading attacks like phishing and improper data modification [21]. Advanced analytics using transformers are applied to detect anomalies, monitor device behavior, and enhance security in real time across complex IoT deployments [22].

There is a suggested cryptosystem that combines homomorphic encryption of integer and floating-point data, using lattice-based constructions and Lotkin functions, with detailed theoretical security evaluations. Privacy and data security are the critical issues in the context of smart grids and home automation due to the high prevalence of smart meters. In order to reduce the risks related to it, one of the architectural frameworks is offered that provides protection to user data, integrates commercial solutions, secures security settings, and eliminates the risks of identity disclosure by keeping sensitive information out of the scope of local networks.

## 7.1. IoT Security

Security solutions shall be rethought and adapted, or even reinvented to suit the specific constraints of these environments, as more heterogeneous the number of IOT nodes and related networks grow [11]. The old guard of security mechanisms that were designed for the computer-focused infrastructures have frequently fallen short when they are transferred and used in decentralized environments, e.g. peer-to-peer systems or large-scale data-driven applications, including smart grids and credit card fraud detection systems [23]. In addition, traditional security models based on assumptions that are hardware- or OS-specific increasingly suffer from vulnerabilities to the newer generation of threats, such as malware, worms (includes propagation exploits), DoS (Denial of Service) attacks and internet-based automated bots. These architectural and deployment changes have created new attack surfaces and vulnerabilities, driving the demand for lightweight, context-aware, and adaptive security paradigms that are suitable to today's IoT landscape.

The rise of what is now more and more commonly referred to as the next generation of World Wide Web, are marked by an enormous increase in diversity and amount of data sources that govern the networking and interconnected devices, sensors, and computer nodes. In contrast to the traditional architecture where only internally and electronically governable computer nodes constitute sources of data; this paradigm shift results in more complex structured and dynamic data streams. Accordingly, the characteristic of saleable information has also changed prescribing much deeper and wider targeted security counter measures [1]. Acknowledging these constraints, particularly the resource- and operationally-constrained environments, this research concentrates on designing and utilizing lightweight cryptography. These algorithms are designed for peer-to-peer models, which seek to reduce device overhead while preserving traffic privacy and alleviating network vulnerably.

## 7.2. Smart Grids

Smart grids use digital technologies for better energy management but pose serious security and privacy issues because of the connectivity among devices (such as smart meters and sensors). Lightweight cryptographic algorithms are needed to protect such devices which have constraints of memory, processing power and energy. The algorithms are evaluated by ASIC and FPGA hardware architecture simulations on a test-board with various parameters found in the practical implementation of SG's devices. The obtained research results have proven the proposed approaches and hardware realizations to be efficient for surveillance of sensors and control of SG working conditions, no matter what kind of implementation technology is used [1].

## 8. Future Directions and Emerging Trends

The Quantum computers are a serious threat to the existing public-key cryptosystems and consequently there is intensive work on post-quantum cryptography. The need for hybrid security architectures, based on the Blockchain, Federated Learning, AI paradigms to secure IoMT-driven healthcare [24], is giving rise to distributed trust management, privacy-preserving model learning and intelligent threat detection in resource-constrained settings with stringent latency limitations. Post-quantum cryptosystems, such as lattice-based and non-lattice-based primitives, are being standardized by NIST where light-weight versions like CLoCK and NTRU promising energy-efficient secure silicon implementations on resource-constrained platforms [25]. Quantum algorithms that are disruptive, such as Shor's have the potential to compromise classical asymmetric schemes and will accelerate the use of quantum-secure cryptography. Application of the proposed algorithms on embedded systems – IoT research, Future work would focus on making sure that these promising security mechanisms are implemented in practice as for instance with validation results performed with synthetic medical data so as to facilitate the privacy-preserving evaluation of these advanced security schemes [26].

## 9. Conclusion and Recommendations

This research exposes the importance of lightweight algorithms for securing low-systems for IoT and embedded systems. The research involved the development a low-power and memory-efficient block cipher that can be designed and implemented to fulfill the radio device s that should satisfy its requirements of computational power as well as energy. Results from hardware and FPGA-based comparisons ensure that the considered lightweight algorithms feature high throughput with low power consumption, thereby being a good candidate for real-time IoT applications.

As a comparison, the best-known features provide that the HIGHT block cipher and hash function SHA-3(cSHAKE) can be implemented on 8-bit microcontroller with approximately 1KB code additional size. FPGA results also suggest that HIGHT is a more area-efficient algorithm than SHA-3 up to 1K LUT architecture, which supports applicability of the proposed lightweight algorithms for hardware-constrained IoT devices. These results validate the feasibility of a lightweight cryptography inclusion in system-on-chip (SoC) designs with a view to achieving secure and cost-effective solutions for IoT and smart city environments [27].

Although traditional cryptographic primitives such as AES are still prevalent, the findings illustrate the importance of lightweight schemes in order to achieve a balance between security and efficiency for resource constrained systems. In future, full security analysis (with respect to birthday and meet-in-the-middle attacks, etc.) against known/related other key and emerging quantum-era attacks of these lightweight cryptographic proposals should be also implemented in order to make the proposed schemes very-robust.

## References

[1] M. Abu-Tair *et al.*, "Towards secure and privacy-preserving IoT enabled smart home: architecture and experimental study," *Sensors*, vol. 20, no. 21, p. 6131, 2020.
[2] S. O. Boufaida, A. Benmachiche, M. Maatallah, and C. Chemam, "An Extensive Examination of Varied Approaches in E-Learning and MOOC Research: A Thorough Overview," in *2024 6th International Conference on Pattern Analysis and Intelligent Systems (PAIS)*, IEEE, 2024, pp. 1–

8. Accessed: Aug. 18, 2024. [Online]. Available: https://ieeexplore.ieee.org/abstract/document/10541129/

[3] S. O. Boufaida, A. Benmachiche, M. Derdour, M. Maatallah, M. S. Kahil, and M. C. Ghanem, "TSA-GRU: A Novel Hybrid Deep Learning Module for Learner Behavior Analytics in MOOCs," *Future Internet*, vol. 17, no. 8, p. 355, 2025.

[4] S. O. Boufaida, A. Benmachiche, A. Bennour, M. Maatallah, M. Derdour, and F. Ghabban, "Enhancing MOOC Course Classification with Convolutional Neural Networks via Lion Algorithm-Based Hyperparameter Tuning," *SN COMPUT. SCI.*, vol. 6, no. 6, p. 707, July 2025, doi: 10.1007/s42979-025-04179-8.

[5] I. Boutabia, A. Benmachiche, A. A. Betouil, and C. Chemam, "A Survey in the Use of Deep Learning Techniques in The Open Classroom Approach," in *2024 6th International Conference on Pattern Analysis and Intelligent Systems (PAIS)*, IEEE, 2024, pp. 1–7. Accessed: Aug. 18, 2024. [Online]. Available: https://ieeexplore.ieee.org/abstract/document/10541268/

[6] I. Boutabia, A. Benmachiche, A. Bennour, A. A. Betouil, M. Derdour, and F. Ghabban, "Hybrid CNN-ViT Model for Student Engagement Detection in Open Classroom Environments," *SN COMPUT. SCI.*, vol. 6, no. 6, p. 684, July 2025, doi: 10.1007/s42979-025-04228-2.

[7] I. Boutabia, A. Benmachiche, A. A. Betouil, M. Boutassetta, and M. Derdour, "A Survey on AI Applications in the Open Classroom Approach," in *2025 International Conference on Networking and Advanced Systems (ICNAS)*, Oct. 2025, pp. 1–7. doi: 10.1109/ICNAS68168.2025.11298090.

[8] D. Abbas, A. Benmachiche, M. Derdour, and B. K. Sedraoui, "Privacy and Security in Decentralized Cyber-Physical Systems: A Survey," in *2025 International Conference on Networking and Advanced Systems (ICNAS)*, Oct. 2025, pp. 1–10. doi: 10.1109/ICNAS68168.2025.11297996.

[9] B. K. Sedraoui, A. Benmachiche, A. Makhlouf, and C. Chemam, "Intrusion Detection with deep learning: A literature review," in *2024 6th International Conference on Pattern Analysis and Intelligent Systems (PAIS)*, IEEE, 2024, pp. 1–8. Accessed: Aug. 18, 2024. [Online]. Available: https://ieeexplore.ieee.org/abstract/document/10541191/

[10] B. K. Sedraoui, A. Benmachiche, A. Bennour, A. Makhlouf, M. Derdour, and F. Ghabban, "LSTM-SWAP: A Hybrid Deep Learning Model for Cheating Detection," *SN COMPUT. SCI.*, vol. 6, no. 7, p. 798, Sept. 2025, doi: 10.1007/s42979-025-04334-1.

[11] M. Abutaha, B. Atawneh, L. Hammouri, and G. Kaddoum, "Secure lightweight cryptosystem for IoT and pervasive computing," *Scientific reports*, vol. 12, no. 1, p. 19649, 2022.

[12] M. A. Al-Shabi, "A survey on symmetric and asymmetric cryptography algorithms in information security," *International Journal of Scientific and Research Publications (IJSRP)*, vol. 9, no. 3, pp. 576–589, 2019.

[13] M. A. H. W. Sohel Rana, A. Azgar, and D. M. A. Kashem, "A survey paper of lightweight block ciphers based on their different design architectures and performance metrics," *Int. J. Comput. Eng. Inf. Technol*, vol. 11, pp. 119–129, 2019.

[14] S. S. Dhanda, B. Singh, and P. Jindal, "Lightweight Cryptography: A Solution to Secure IoT," *Wireless Pers Commun*, vol. 112, no. 3, pp. 1947–1980, June 2020, doi: 10.1007/s11277-020-07134-3.

[15] M. Rana, Q. Mamun, and R. Islam, "Lightweight cryptography in IoT networks: A survey," *Future Generation Computer Systems*, vol. 129, pp. 77–89, 2022.

[16] A. Benmachiche, K. Rais, and H. Slimi, "Real-Time Machine Learning for Embedded Anomaly Detection," Dec. 22, 2025, *arXiv*: arXiv:2512.19383. doi: 10.48550/arXiv.2512.19383.

[17] M. Rana, Q. Mamun, and R. Islam, "Current Lightweight Cryptography Protocols in Smart City IoT Networks: A Survey," Oct. 02, 2020, *arXiv*: arXiv:2010.00852. doi: 10.48550/arXiv.2010.00852.

[18] A. Shahverdi, "Lightweight Cryptography Meets Threshold Implementation: A Case Study for Simon," *Master's Thesis. Worcester Polytechnic Institute*, 2015, Accessed: Jan. 04, 2026. [Online]. Available: https://digital.wpi.edu/downloads/qb98mf61z?utm_source=chatgpt.com

[19]   S. B. Sadkhan and A. O. Salman, "A survey on lightweight-cryptography status and future challenges," in *2018 International Conference on Advance of Sustainable Engineering and its Application (ICASEA)*, IEEE, 2018, pp. 105–108. Accessed: Jan. 06, 2026. [Online]. Available: https://ieeexplore.ieee.org/abstract/document/8370965/

[20]   N. A. Gunathilake, W. J. Buchanan, and R. Asif, "Next generation lightweight cryptography for smart IoT devices:: implementation, challenges and applications," in *2019 IEEE 5th World Forum on Internet of Things (WF-IoT)*, IEEE, 2019, pp. 707–710. Accessed: Jan. 06, 2026. [Online]. Available: https://ieeexplore.ieee.org/abstract/document/8767250/

[21]   B. K. Sedraoui, A. Benmachiche, A. Makhlouf, D. Abbas, and M. Derdour, "Cybersecurity in E-Learning: A Literature Review on Phishing Detection Using ML and DL Techniques," in *2025 International Conference on Networking and Advanced Systems (ICNAS)*, Oct. 2025, pp. 1–10. doi: 10.1109/ICNAS68168.2025.11298114.

[22]   K. Rais, M. Amroune, and M. Y. Haouam, "CAT-VAE: A Cross-Attention Transformer-Enhanced Variational Autoencoder for Improved Image Synthesis".

[23]   I. Soualmia, S. Maalem, A. Benmachiche, K. Rais, and M. Derdour, "Comparative Survey of AI-Driven Credit Card Fraud Detection: Machine Learning, Deep Learning and Hybrid Systems," in *2025 International Conference on Networking and Advanced Systems (ICNAS)*, Oct. 2025, pp. 1–9. doi: 10.1109/ICNAS68168.2025.11298125.

[24]   R. Mounira, M. Majda, B. Abdelmadjid, B. S. Oumaima, and M. Derdour, "A Comprehensive Survey on Blockchain, Federated Learning, and AI for Securing the Internet of Medical Things," in *2025 International Conference on Networking and Advanced Systems (ICNAS)*, Oct. 2025, pp. 1–8. doi: 10.1109/ICNAS68168.2025.11298047.

[25]   N. A. Gunathilake, A. Al-Dubai, W. J. Buchanan, and O. Lo, "Electromagnetic Analysis of an Ultra-Lightweight Cipher: PRESENT," June 29, 2021. doi: 10.1521/csit.2021.110915.

[26]   K. Rais, M. Amroune, M. Y. Haouam, and A. Benmachiche, "Evaluating Class Integrity in GAN-Generated Synthetic Medical Datasets," in *2025 International Conference on Networking and Advanced Systems (ICNAS)*, Oct. 2025, pp. 1–8. doi: 10.1109/ICNAS68168.2025.11298012.

[27]   M. Boutassetta, A. Makhlouf, N. Messaoudi, A. Benmachiche, I. Boutabia, and M. Derdour, "Cyberattack Detection in Smart Cities Using AI: A literature review," in *2025 International Conference on Networking and Advanced Systems (ICNAS)*, Oct. 2025, pp. 1–9. doi: 10.1109/ICNAS68168.2025.11298103.